\documentclass{aastex631}

\usepackage{CJK}

\shorttitle{Exploring the FRB-BNS link with space GW observatories}
\shortauthors{Yin et al.}
\graphicspath{{./}{figures/}}

\begin{document}
\begin{CJK*}{UTF8}{gbsn}

\title{Exploring the link between FRBs and binary neutron stars Origins with Space-borne Gravitational Wave Observations}
\author{Yu-xuan Yin(尹宇轩)}
\affiliation{MOE Key Laboratory of TianQin Mission \\ TianQin Research Center for Gravitational Physics \& School of Physics and Astronomy \\ Frontiers Science Center for TianQin Gravitational Wave Research Center of CNSA \\ Sun Yat-sen University (Zhuhai Campus), Zhuhai 519082, China}


\author{En-kun Li(李恩坤)}
\affiliation{MOE Key Laboratory of TianQin Mission \\ TianQin Research Center for Gravitational Physics \& School of Physics and Astronomy \\ Frontiers Science Center for TianQin Gravitational Wave Research Center of CNSA \\ Sun Yat-sen University (Zhuhai Campus), Zhuhai 519082, China}

\author{Bing Zhang(张冰)}
\affiliation{Nevada Center for Astrophysics, University of Nevada, Las Vegas, NV 89154, USA}
\affiliation{Department of Physics and Astronomy, University of Nevada, Las Vegas, NV 89154, USA}

\author{Yi-Ming Hu(胡一鸣)}
\email{huyiming@sysu.edu.cn}  
\affiliation{MOE Key Laboratory of TianQin Mission \\ TianQin Research Center for Gravitational Physics \& School of Physics and Astronomy \\ Frontiers Science Center for TianQin Gravitational Wave Research Center of CNSA \\ Sun Yat-sen University (Zhuhai Campus), Zhuhai 519082, China}

\begin{abstract}
The origin of repeating Fast Radio Bursts (FRBs) is an open question, with observations suggesting that at least some are associated with old stellar populations. 
It has been proposed that some repeating FRBs may be produced by interactions of the binary neutron star magnetospheres decades to centuries before the coalescence. 
These systems would also emit centi-Hertz gravitational waves during this period, which can be detectable by space-borne gravitational wave detectors.
We explore the prospects of using current and future space-borne gravitational wave detectors, such as TianQin, LISA, and DECIGO, to test this FRB formation hypothesis.
Focusing on nearby galaxies like M81, which hosts a repeating FRB source in a globular cluster, we calculate the detection capabilities for binary neutron star systems. 
Our analysis reveals that while missions like TianQin and LISA face limitations in horizon distance, changing detector pointing direction could significantly enhance detection probabilities. Considering the chance of a Milky Way-like galaxy coincidentally containing a BNS within 100 years before merger is only $3\times10^{-5}$ to $5\times10^{-3}$, if a signal is detected originating from M81, we can establish the link between FRB and binary neutron stars with a significance level of at least 2.81$\sigma$, or a Bayes factor of $4\times10^6 - 7\times10^8$ / $5\times10^2 - 10^5$ against the background model with optimistic/realistic assumptions.
Next-generation detectors such as DECIGO offer enhanced capabilities and should easily detect these systems in M81 and beyond. Our work highlights the critical role of space-borne gravitational wave missions in unraveling FRB origins.

\end{abstract}



\section{Introduction} \label{sec:intro}

Fast radio Bursts (FRBs) are millisecond-duration radio bursts from cosmological distances \citep{lorimer2007,thornton13}. Initially detected as one-off events, later observations suggested that at least some, probably most, FRBs are repeating sources \citep{spitler2016,chime-repeater1,chime-repeater2}. Thanks to the extensive observations using powerful radio telescopes such as 
Parkes, 
Canadian Hydrogen Intensity Mapping Experiment (CHIME) \citep{andersen2019chime,amiri2021first}, Australian Square Kilometre Array Pathfinder (ASKAP) \citep{johnston2008science,Kumar_2019}, and Five-Hundred-Meter Aperture Spherical Radio Telescope (FAST) \citep{nan2006five,luo2020,lid21,niu2022repeating,xu22}, our understanding of FRBs has been greatly advanced \citep{zhang2023physics}. Albeit it was confirmed that magnetars can make FRBs as observed as FRB 20200428D from the Milky Way Galaxy \citep{FRB200428-CHIME,FRB200428-STARE2,lin00,FRB200428-Integral,FRB200428-HXMT}, the origin of the cosmological FRBs is still not settled. Even though some active repeaters are consistent with being powered by young magnetars from star-forming regions \citep{tendulkar17,bhardwaj24}, at least some FRBs show evidence of delay with respect to star formation \citep{zhangzhang22,sharma24}, and therefore, are potentially associated with an old stellar population. In particular, the repeating rFRB 20200120E was found to be associated with a globular cluster of the nearby galaxy M81 \citep{bhardwaj21,kirsten2022repeating,zhangsb24}, suggesting a distinct type of FRB source or a distinct channel of forming magnetars. Some (not all) FRB sources show significant offset from the host galaxies \citep{marcote2020,bannister2019single}, which could be consistent with the short GRBs that are believed to be produced by binary neutron star (BNS) mergers \citep{grindlay2006short,fong10,lizhang20}. 


One possible model to account for repeating FRBs in old stellar populations was proposed by \cite{zhang2020}, who suggested that magnetospheric interactions between two neutron stars decades to centuries before the coalescence could induce magnetic reconnections that might launch narrowly beamed outflows to power repeated bursts. Such BNS systems are also gravitational wave emitters. \cite{zhang2020} suggested that detecting long-term gravitational waves from repeating FRB sources would provide definite proof of such an FRB source model.   


About 20 BNSs have been discovered through radio pulsar surveys \citep{andrews2019double}. In GW surveys, LIGO \citep{aasi2015advanced} and Vrigo \citep{acernese2014advanced} have discovered 2 BNS merger events, GW170817 \citep{abbott2017gw170817} and GW190425 \citep{abbott2020gw190425}. From these two observational results, the estimated BNS merger rate is between 10 and 1700 $\mathrm{Gpc^{-3} yr^{-1}}$ \citep{abbott2023population}. 
Several works have estimated the detectable BNS number in the Milky Way. \cite{lau2020detecting} found that for a 4-year observation, LISA can detect 35 BNSs, 33 in our galaxy and 2 in Large Magellanic Cloud and Small Magellanic Cloud, with the threshold of signal-to-noise ratio (SNR) greater than 8. \cite{feng2023multimessenger} simulated the BNS population in the Galactic disk and found that the average number of BNSs detectable by TianQin, LISA, and TianQin+LISA are 217, 368, and 429, respectively. However, these studies have mostly focused on the detection of BNSs in the Milky Way and have not considered the possibility of repeating FRB sources (which are all extragalactic sources) as possible BNS systems. In particular, it is interesting to investigate the detectability of some repeating FRB sources by TianQin, LISA, and other detectors, especially for nearby sources such as rFRB 20200120E in M81. 

In this work, we specifically focus on the detection of BNSs within the Local Group, with a pivotal emphasis on M81, where the promising repeating FRB source rFRB20200120E resides in a globular cluster at a distance of 3.6 Mpc \citep{bhardwaj21,kirsten2022repeating,zhangsb24}. Unlike previous studies that have predominantly analyzed BNS detections within the Milky Way, our primary innovation lies in examining the capability of detectors, exemplified by TianQin, LISA, and DECIGO, to identify such systems in external galaxies, especially M81.
Regarding these three detectors, LISA is scheduled to commence operations in the 2030s \citep{LISA:2017pwj}, and TianQin is expected to be put into operation around 2035 \citep{TianQin:2020hid}. For DECIGO, it is planned to launch B-DECIGO as a precursor to DECIGO during the 2030s \citep{Kawamura:2020pcg}. 
By doing so, we aim to shed light on the physical origin of some repeating FRB sources and underscore the unique role that space-borne gravitational wave detectors can play in unraveling these cosmic mysteries.

The remainder of this paper is structured as follows. Section \ref{sec:method} briefly introduces the gravitational wave (GW) waveform emitted by BNS and discusses the methodology for detecting such signals. Section \ref{sec:result} presents the detection capabilities of the TianQin detector for BNS systems in various galaxies, with a primary focus on M81. This section also includes a comparison of the detection capabilities of LISA and DECIGO for BNS systems. Finally, Section \ref{sec:conclu} summarizes the findings and concludes the paper.

\section{Methodology} \label{sec:method}
For an isolated BNS system, under quasi-circular orbit, GW signals are given by: 

\begin{equation}
    h_{+}(t)=\mathcal{A}(1+\cos^2\iota)\cos\Phi(t),
\end{equation}
\begin{equation}
    h_{\times}(t)=2\mathcal{A}\cos\iota\sin\Phi(t),
\end{equation}
where the amplitude is $\mathcal{A}=\frac{2(G\mathcal{M})^{5/3}}{c^4d}(\pi f)^{2/3}$. $\mathcal M$ is chirp mass, $d$ is luminosity distance of the source, and $\Phi(t) = \Phi_c -2\pi \int_t^{t_c}f {\rm d}t$ with $\Phi_c = \Phi(t_c)$. Considering a typical BNS system with equal masses $m_1=m_2=1.4M_{\odot}$ and a time remaining before merger $t_c-t=100 $ years, the GW frequency is calculated to be $f_{GW}=3.6\times 10^{-2}$Hz. The rate of change of this frequency, $\dot{f}_{GW}=1.3\times 10^{-4}$ Hz per year. Given that the designed operational lifetime of the detector is several years, and the change in frequency will be two orders of magnitude smaller than the frequency, out of simplicity we assume the frequency $f$ to be constant during our observational period so that $\Phi(t) = \Phi_c -2\pi(t_c-t)$.

The signal detected by TianQin or LISA can be formulated as described by (\cite{cutler1998angular})
\begin{equation}
    h_{\alpha}=h_+(t)F_{\alpha}^+(t)+h_{\times}(t)F_{\alpha}^{\times}(t), \alpha=\uppercase\expandafter{\romannumeral 1}, \uppercase\expandafter{\romannumeral 2}.
\end{equation}
Here, $F$ represents the antenna response function specific to different detectors. We can calculate the optimal SNR as the inner product of the signal $h$ with itself:
\begin{equation}
    \rho^2=(h|h)=\frac{2}{S_n(f_0)}\int_0^T h^2\mathrm dt.
\end{equation}
$S_n(f)$, which is characterized by the acceleration noise $S_a$ and positional noise $S_x$, represents the sensitivity curve of detectors, and $T$ is observation time. Notice that TianQin is expected to operate with a 3 months on/3 months off scheme, so we take the observation time to be half the operation time of five years. 

We use the Fisher information metrics (FIM) to calculate the parameter estimation precision
\begin{equation}
    \Gamma_{ij}\equiv\left (\frac{\partial h(\mathbf{\theta})}{\partial \theta_i} \middle| \frac{\partial h(\mathbf{\theta})}{\partial \theta_j}\right)
\end{equation}
The error of parameter $\theta_i$ can be estimated as $\Delta \theta_i=\sqrt{\Sigma_{ii}}$, where $\Sigma=\Gamma^{-1}$. For sky localization of the source $(\theta,\phi)$, the sky localization uncertainty is
\begin{equation}
    \Delta\Omega=2\pi|\sin\theta|\sqrt{\Sigma_{\theta\theta}\Sigma_{\phi\phi}-\Sigma_{\theta\phi}^2}.
\end{equation}

We compute the Bayes factor to compare the support for the rFRB-BNS association model($M_1$) against the null hypothesis of a coincidental correlation($M_0$). The Bayes factor is defined as the ratio of the marginal likelihoods of the two models given the observed data $D$.
\begin{equation}
    BF=\frac{p(D|M_1)}{p(D|M_0)}
\end{equation}

\section{Results} \label{sec:result}
\subsection{Detection Capabilities of TianQin and LISA}

\begin{figure}[ht!]
\centering
\includegraphics[width=0.6\textwidth]{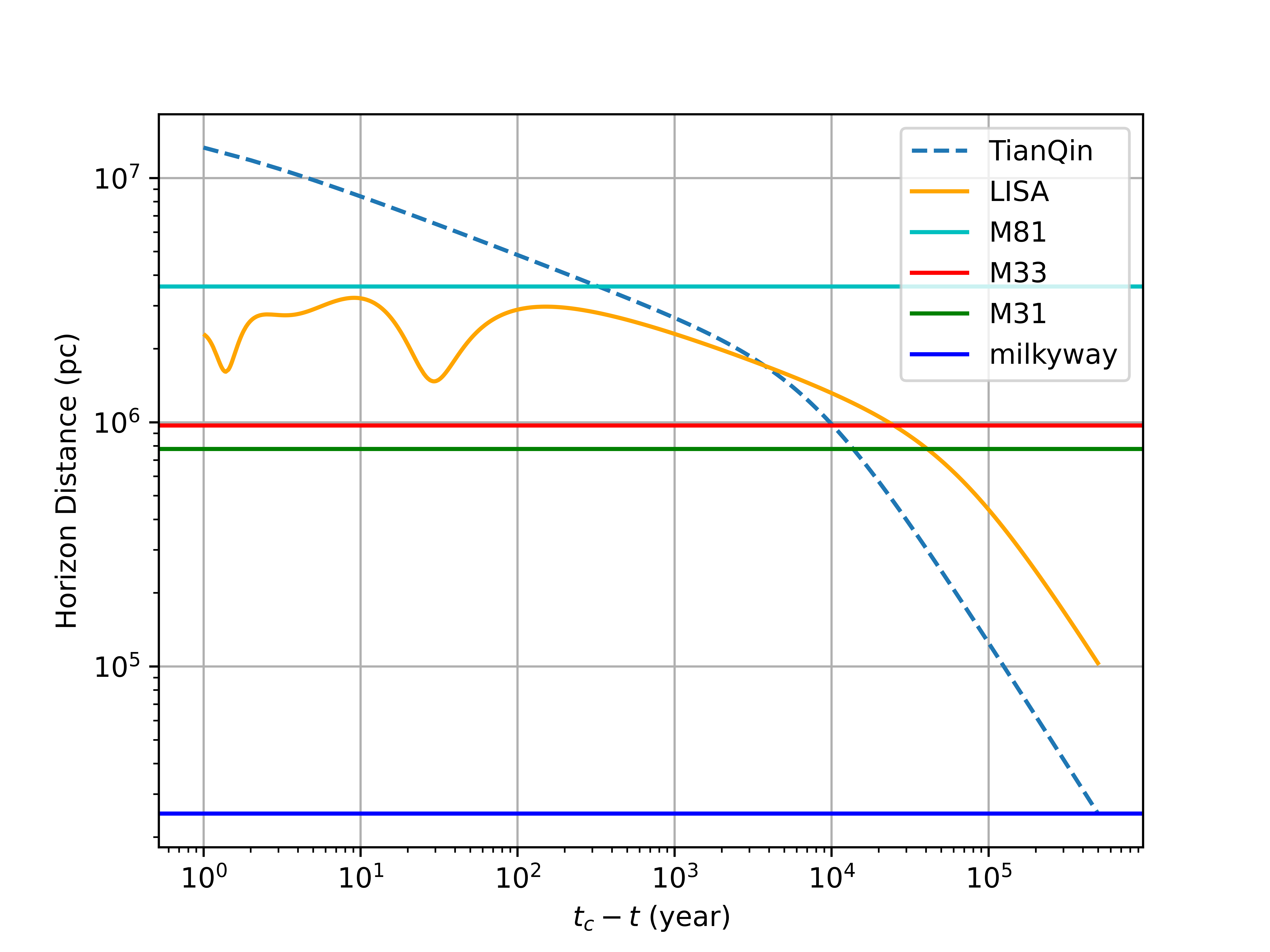}
\caption{
The TianQin/LISA horizon distance for BNS systems at various evolution stages of coalescence, as denoted by $t_c-t$. TianQin/LISA could observe sources beneath the dark blue dashed curve/orange solid curve with $\rho>7$. The cyan/green/red/blue lines indicate the distances to M81/M33/M31/opposite edge of the Milky Way, respectively.\label{fig:fig2}}
\end{figure}

We first calculate the horizon distances to assess the detection capabilities of TianQin and LISA for BNS systems.
The horizon distance is the furthest distance that a source is still detectable, so all geometrical configurations are optimal. 
Considering two neutron stars, each with a mass of $1.4M_{\odot}$, and setting the detection threshold to be $\rho_0=7$, we show the horizon distances for TianQin at various coalescence times in Figure \ref{fig:fig2}.
If the source is located at the pole of the TianQin constellation, and the binary angular momentum aligns with the line-of-sight, then TianQin could detect BNS systems in M81, M33, M31, and the opposite edge of the Milky Way up to 351 years, 10,185 years, 13,721 years, and 500,000 years before the merger, respectively.

The results of LISA are shown in orange.
The orbital plane of LISA is not constant, so there is no physical counterpart of the optimal location. 
For the sake of a fair comparison, we ignore the antenna response and compute the SNR over the same operation duration of 5 years, which is longer than the nominal 4-year mission time.
Since LISA has better sensitivity in lower frequencies, for BNS systems that merge thousands of years later, LISA has a further horizon distance.
However, the worse sensitivity in higher frequencies indicates that LISA can not make detections if the BNS is located at the distance of M81. 

\begin{figure}[ht!]
\centering
\includegraphics[width=0.6\textwidth]{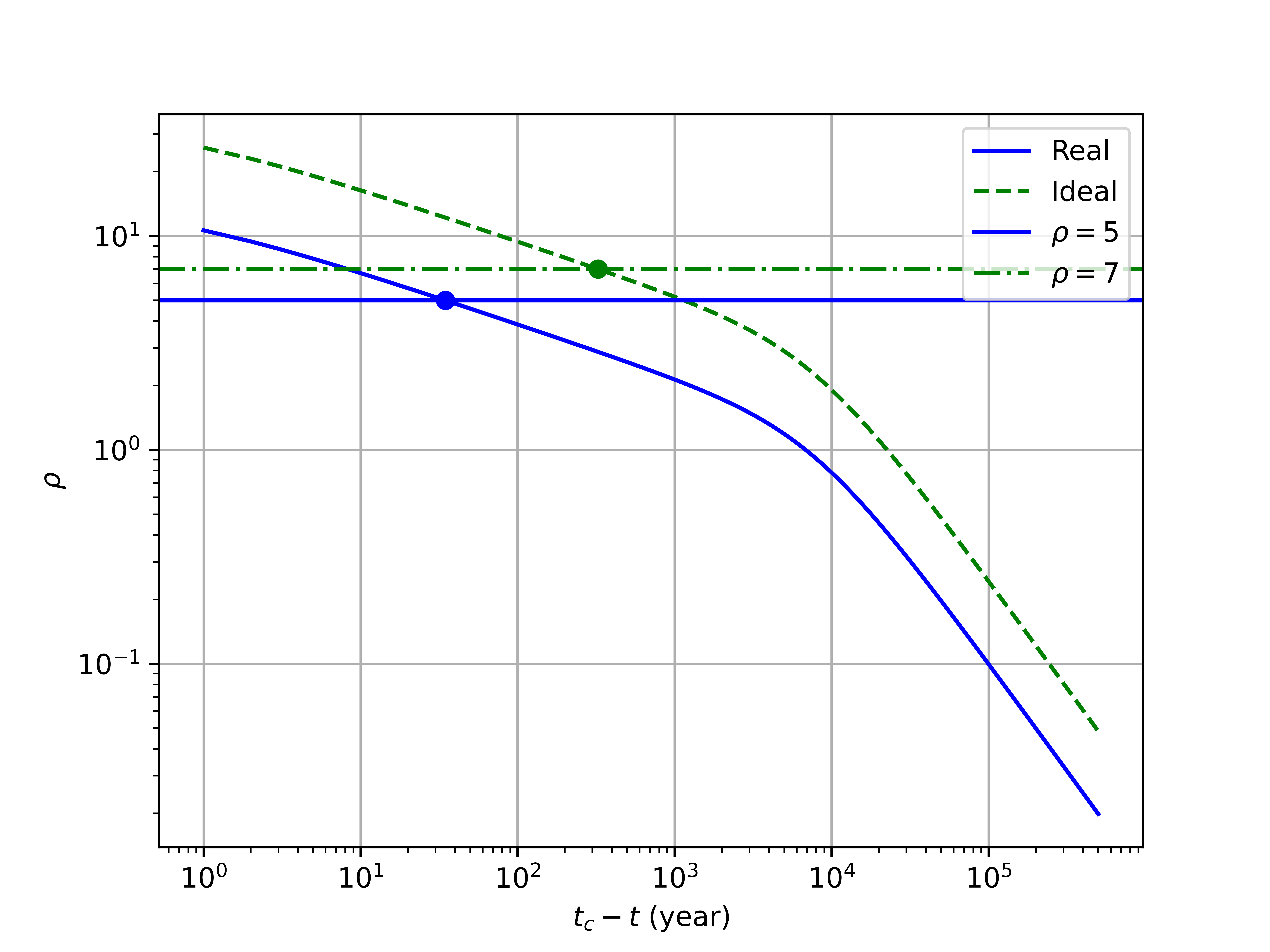}
\caption{Evolution of the SNR of a BNS in M81. The blue solid curve represents the SNR results that account for the antenna response effect, whereas the green dashed curve shows the results without considering the antenna response effect. The green dash-dot line and the blue solid line denote the SNR thresholds of 7 and 5, respectively.}\label{fig:fig3}
\end{figure}

TianQin has a weaker response to a general source in more realistic cases.
The response to a source located in the TianQin constellation pole can reach 100\%, but this response can drop if the source is located anywhere else.
On the other hand, if we consider the exact location of the source during the search, then it will become a directed search instead of a blind search, in which one can adopt a lower detection threshold. 
We illustrate the effect of antenna response in Figure \ref{fig:fig3}, for a BNS system in M81, taking the geometric information into account led to a factor of 3 drop in SNR. 
When considering a detection threshold of $\rho_0=5$, the observation window for a BNS in M81 shrinks to 36 years before the merger. 

\begin{figure}[ht!]
\centering
\includegraphics[width=0.6\textwidth]{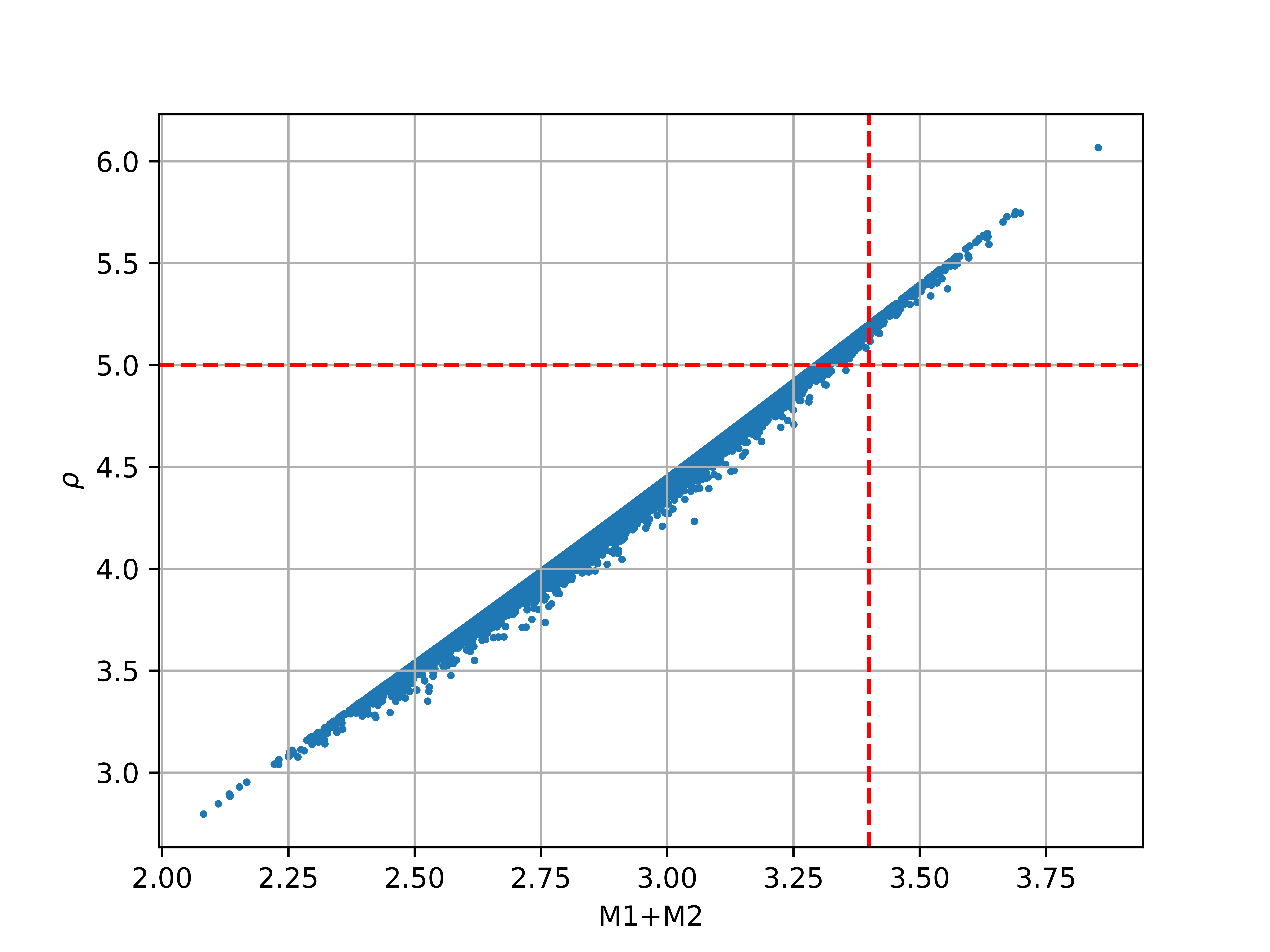}
\caption{SNR distribution for BNS systems assuming a double Gaussian mass distribution. Considering a BNS at M81 that will merge in 80 years, there's a 1.5\% probability it will be detectable, while a GW190425-like BNS system with a total mass of $3.4M_{\odot}$ will definitely be detectable.  \label{fig:fig4}}
\end{figure}

We then consider a more realistic mass distribution of the BNS systems.
Following \cite{farrow2019}, we consider a double Gaussian distribution for the BNS masses with the following parameters: $\mu_1=1.34M_{\odot}$, $\mu_2=1.47M_{\odot}$; $\sigma_1=0.02$, $\sigma=0.15$; $\alpha=0.68$. 
For BNS systems that will merge in 80 years, we generate a total of $5\times 10^4$ BNS systems and calculate the SNR.
As shown in Figure \ref{fig:fig4}, the SNR strongly correlates with the total mass.
Under the double Gaussian distribution, about 1.5\% of the BNS systems will exceed the SNR threshold of 5 and become detectable.
Although the ratio is low, we notice that if we consider a GW190425-like BNS system with a total mass of $3.4M_{\odot}$ \citep{romero2020}, then its signal is detectable.

We also calculate the Fisher matrix to analyze the sky localization ability of the two detectors. For BNS systems in M81, the sky localization uncertainty $\Delta \Omega$ of TianQin and LISA is approximately 0.45 deg$^2$ and 0.23 deg$^2$, respectively. At the distance of M81, this corresponds to a position uncertainty of approximately 20 kpc. Considering that the M81 is 30 kpc across, this is sufficient to determine whether the signal is located within the M81 galaxy. 
However, even with a worse sky localization ability, the BNS detections through space-borne GW missions can still establish the link.
First of all, the galaxy number density is sparse enough that if a BNS is identified in the rough sky region and distance, we can associate its host galaxy confidently.
Secondly, the expected merger rate for BNSs in a Milky Way-like galaxy is so low (between 10 and 1700 $\mathrm{Gpc^{-3} yr^{-1}}$ \citep{abbott2023population}), that the chance of it coincidentally containing a BNS within 100 years before merger is only $3\times10^{-5}-5\times10^{-3}$.
This means that once a BNS system is detected in M81, the rFRB-BNS link can be established with a significance of 2.81$\sigma$ to 4.17$\sigma$. 

By comparing the hypotheses of a rFRB-BNS link origin versus an coincidental correlation, we calculate Bayes factors of approximately $4\times10^6$ (corresponding to a merger rate of 1700$\mathrm{Gpc^{-3} yr^{-1}}$) to $7\times10^8$ (corresponding to a merger rate of 10$\mathrm{Gpc^{-3} yr^{-1}}$) for TianQin and LISA. These values are derived under the idealized assumption of uniform event distribution across the sky. However, in reality, we expect that the events only occur within galaxies. In this case, the Bayes factor is approximately $5\times10^{2}$ to $10^5$, which is still strong enough to provide compelling evidence in support of the rFRB-BNS association model. The Bayes factor calculation indicates that even with moderate sky localization uncertainty, a confident identification for whether the signal has the same origin as rFRBs can be achieved.

\subsection{Effects of the pointing of TianQin}
From the previous study, we realize that the geometric effect plays a significant role, and the current TianQin design does not guarantee detection for a BNS system in M81 if it merges in a century. 
We then examine how shifting the pointing of the TianQin constellation can affect the outcome. 
By assuming a 1.4 $M_\odot$ equal mass BNS and a merger time of 100 years, we simulate catalogs of BNS systems with random orientations and calculate the SNR.
We consider three pointing direction scenarios: 1. J0806, which is the nominal design; 2. M81, which is the host galaxy of the rFRB 20200120E; 3. M81 projection on the ecliptic plane. 
The final case is a compromise between science output and practical consideration: previous studies reveal that if the orbital plane of TianQin is perpendicular to the ecliptic plane, it will achieve optimal orbital and thermal stability \citep{Ye:2024dca,Chen:2021dzg}. 

The probabilities of detecting BNS in different scenarios are presented in Table \ref{tab:2}. 
In the three scenarios, the probabilities are 0\%, 55\%, and 41\% 
Keeping the nominal pointing to J0806, no matter what orientation the BNS is, as long as it is located in M81 and merges in 100 years, it will not be detectable for TianQin.
Pointing TianQin at M81 will significantly boost the chance.
Limiting the pointing on the ecliptic plane leads to a slight decrease, but there is still a good chance of observation.

\begin{deluxetable*}{cccc} [h]
\tablecaption{Probability of detecting a BNS system from M81 under different TianQin paintings. 
We fix the component masses to be 1.4 $M_\odot$ and the merger time to be 100 years. 
For a randomly oriented BNS system, the chances of TianQin detection are 0\%, 55\%, and 41\%, for the pointing of J0806/M81/M81 projection on the ecliptic plane, respectively. \label{tab:2}}  
\tablewidth{0pt} 
\tablehead{  
\colhead{Pointing Direction} & \colhead{Ecliptic Longitude (deg)} & \colhead{Ecliptic Latitude (deg)} & \colhead{Probabilities of Detecting BNS (\%)}  
}   
\startdata  
J0806 & 120.5 & -4.7 & 0 \\  
M81 & 50.79 & 56.42 & 55 \\  
M81 projection on the ecliptic plane & 50.79 & 0 & 41 \\  
\enddata  
\end{deluxetable*}

\subsection{Detection Prospect with DECIGO}
Next-generation detectors such as DECIGO \citep{kawamura2011japanese} are expected to have enhanced detection capability than TianQin and LISA. 
DECIGO will contain four units of triangle-like detectors. The noise spectrum is given by \citep{yagi2011detector}
\begin{equation}
    S_h^{\mathrm{DECIGO}}(f)=7.05\times10^{-48}\left[1+\left(\frac{f}{7.36 {\rm Hz}}\right)^2\right]+4.8\times10^{-51}\left(\frac{f}{1 \mathrm{Hz}}\right)^{-4}\frac{1}{1+\left(\frac{f}{7.36{\rm Hz}}\right)^2}+5.33\times10^{-52}\left(\frac{f}{1 \mathrm{Hz}}\right)^{-4} \mathrm{Hz^{-1}}.
\end{equation}

The anticipated noise spectrum for DECIGO is significantly better than TianQin and LISA.
Our calculation indicates that over one year of observation, DECIGO can observe a BNS in M81 with an SNR of 1466, assuming a 1.4-1.4 $M_\odot$ BNS that merges in a century. 
This SNR increases to 3235 if an observation time of five years is assumed. 
Notice that with five years, for a 1.4-1.4 $M_\odot$ BNS that merges in a century, the horizon distance of DECIGO can reach the level of 1.66 Gpc.
This distance already encloses many observed FRB repeaters.
It is highly likely that the pre-merger FRB hypothesis \citep{zhang2020} can be adequately tested in the DECIGO era. We also calculated the sky localization uncertainty using FIM for DECIGO and $\Delta\Omega_{\rm DECIGO}\simeq4\times10^{-3}$deg$^2$, note that it is a sky average result. In the future, DECIGO could pinpoint the BNS in a circle with a projected radius of about 3 kpc. This enhanced localization precision would increase the Bayes factor up to four orders of magnitude, reaching values between $10^{10}$ and $10^{12}$ under ideal case(uniform event distribution), $10^4$ to $10^6$ when accounting for the events are confined in galaxies. Although this is not enough to pinpoint the host cluster, the improved sky localization can greatly boost the statistical significance of the rFRB-BNS association through a Bayes factor-style argument using volumetric priors. 

\section{Conclusion}\label{sec:conclu}
This study explores the possibility of detecting BNS systems that could potentially serve as the sources of some repeating FRBs \citep{zhang2020} using the space-borne gravitational wave detector.
By using TianQin, LISA, and DECIGO as illustrative examples, we've comprehensively analyzed the detection prospects across different cosmic distances and under diverse observational conditions. 
Our findings highlight the significant yet variable detection capabilities among these detectors. 

For TianQin, it has a significant detection capability for BNS systems up to several hundred years before their mergers, depending on their distances. 
Specifically, if considering a face-on binary located at the pole of the TianQin constellation (J0806), TianQin could detect such systems 351/13,721/500,000 years before the merger, if the source is as far as M81/M31/opposite edge of the Milky Way. 
For a source located in M81, TianQin can only detect 36 years before the merger. We also demonstrate that by adjusting TianQin's pointing direction towards M81, the detection probability can be significantly increased from 0\% (when pointing towards J0806) to 41\% when the ecliptic longitude is aligned with M81, and further to 55\% when TianQin is pointed directly towards M81.

For LISA, it struggles to detect BNS systems at the distance of M81, due to the relatively lower sensitivity. FIM calculation result shows the sky localization uncertainty for TianQin/LISA is smaller than 0.5deg$^2$. If future TianQin/LISA detects a BNS signal from M81, then the association of rFRB and BNS can be established with a significance of 2.81$\sigma$ to 4.17$\sigma$. For TianQin and LISA, Bayes factors for rFRB-BNS associations range from $4\times10^6$ to $7\times10^8$ in under ideal assumption of uniform event distribution, dropping to $5\times10^2$ to $10^5$ when accounting for the events are confined in galaxies.
Next-generation detectors like DECIGO are expected to have significantly enhanced detection capabilities compared to TianQin and LISA, making them promising candidates for detecting BNS systems that could be the sources of repeating FRBs. DECIGO has lower sky localization uncertainty with $4\times10^{-3}$deg$^2$ and can boost the Bayes factor by up to four orders of magnitude ($10^{10}-10^{12}$ ideally and $10^4-10^6$ realistically).

In summary, our work underscores the potential of using space-borne GW detectors in exploring the possible origin of repeating FRBs by detecting pre-merger BNS systems that emit continuous GWs.

\section*{Acknowledgments}
This work has been supported by the National Key Research and Development Program of China (No. 2023YFC2206701), the Natural Science Foundation of China (Grants  No.  12173104, No. 12261131504), and the science research grants from the China Manned Space Project (CMS-CSST-2025-A13).

\bibliography{main}{}
\bibliographystyle{aasjournal}

\end{CJK*}
\end{document}